\documentclass[%
reprint,
showpacs,
bibnotes,
amsmath,amssymb,
aps,
prb,
]{revtex4-1}

\usepackage{graphicx}% Include figure files
\usepackage{dcolumn}% Align table columns on decimal point
\usepackage{bm}% bold math

\usepackage{upgreek}
\usepackage{epsfig}
\usepackage{mathrsfs}
\usepackage{amssymb}
\usepackage{amsbsy}
\usepackage{color}
\usepackage{cancel}
\usepackage{epsf}
\newcommand{\bcen}{\begin{center}}
\newcommand{\ecen}{\end{center}}
\newcommand{\btab}{\begin{tabular}}
\newcommand{\etab}{\end{tabular}}
\newcommand{\bdes}{\begin{description}}
\newcommand{\edes}{\end{description}}

\newcommand{\beq}{\begin{equation}}
\newcommand{\eeq}{\end{equation}}
\newcommand{\bea}{\begin{eqnarray}}
\newcommand{\eea}{\end{eqnarray}}
\newcommand{\non}{\nonumber}

\newcommand{\half}{\frac{1}{2}}
\newcommand{\bary}{\begin{array}}
\newcommand{\eary}{\end{array}}
\newcommand{\benum}{\begin{enumerate}}
\newcommand{\eenum}{\end{enumerate}}
\newcommand{\bitem}{\begin{itemize}}
\newcommand{\eitem}{\end{itemize}}

\newcommand{\btau}{\mbox{\boldmath $ \tau $}}
\newcommand{\blam}{{\boldsymbol{\lambda}}}

\newcommand{\bOne}{{\boldsymbol 1}}
\newcommand{\be} { \mbox{\boldmath $e$}}

\newcommand{\bk} { \bm{k} }

\newcommand{\bp} { \bm{p} }

\newcommand{\br} { \boldsymbol{r}}

\newcommand{\D}[1]{\mbox{d}{#1}}

\newcommand{\ket}[1]{| #1 \rangle}

\newcommand{\prn}[1] {(\ref{#1})}

\newcommand{\fig}[1]{fig.~\ref{#1}}
\newcommand{\Fig}[1]{Fig.~\ref{#1}}

\newcommand{\citebyname}[1]{\citeauthor{#1}\cite{#1}}

\newcommand{\myfigwidth}{0.99\columnwidth}

\newcommand{\asc}{a_{sc}}
\newcommand{\as}{a_{s}}
\newcommand{\cF}{{\cal F} }
\newcommand{\ie}{{i.e., } }

\begin{document}

%\preprint{}

%\preprint{}
\title{How does a synthetic non-Abelian gauge field influence the bound
 states of two spin-$\half$ fermions?}

\author{Jayantha P. Vyasanakere}
\email{jayantha@physics.iisc.ernet.in}
\author{Vijay B. Shenoy}
\email{shenoy@physics.iisc.ernet.in}
% \affiliation{Materials Research Centre, Indian Institute of Science, Bangalore 560 012}%Lines break automatically or can be forced with \\
%\thanks{VBS, {\tt shenoy@physics.iisc.ernet.in}}%
%\email{shenoy@physics.iisc.ernet.in}
\affiliation{Centre for Condensed Matter Theory, Department of Physics, Indian Institute of Science, Bangalore 560 012}
%\affiliation{Centre for Condensed Matter Theory, Indian Institute of Science, Bangalore 560 012}


\date{\today}

\begin{abstract}
We study the bound states of two spin-$\half$ fermions interacting via a
contact attraction (characterized by a scattering length) in the
singlet channel in $3D$ space in presence of a uniform non-Abelian
gauge field. The configuration of the gauge field that generates a
Rashba type spin-orbit interaction is described by three coupling
parameters $(\lambda_x, \lambda_y, \lambda_z)$. For a generic gauge
field configuration, the critical scattering length required for the
formation of a bound state is {\em negative}, \ie shifts to the
``BCS side'' of the resonance.  Interestingly, we find that there are
special high-symmetry configurations (e.g., $\lambda_x = \lambda_y =
\lambda_z$) for which there is a two body bound state for {\em any}
scattering length however small and negative. Remarkably, the bound state wave functions obtained for high-symmetry configurations have nematic spin structure similar to those found in liquid $^3$He.  Our results show that
the BCS-BEC crossover is drastically affected by the presence of a
non-Abelian gauge field. We discuss possible experimental signatures
of our findings both at high and low temperatures.
\end{abstract}


\maketitle

\section{Introduction}

Quantum emulation experiments with cold quantum
gases\cite{Ketterle2008,Bloch2008,Giorgini2008} and optical
lattices hold the promise of providing clues to understanding many
outstanding issues of quantum condensed matter physics such as high
temperature superconductivity, quantum hall effect, etc, and the high
energy physics of strongly coupled gauge
theories.\cite{Maeda2009} While this has led to a flurry
of activity, many experimental challenges remain in the way of
redemption of this promise. Particular among them are the problem of
entropy removal and the creation of magnetic (gauge)
fields.

Realization of magnetic fields has been achieved by rotation
\cite{Cooper2008}, however, attaining magnetic fields corresponding to
quantum hall regimes has serious experimental challenges.  There have
been many theoretical suggestions for the generation of artificial gauge
fields\cite{Jaksch2003,Osterloh2005,Ruseckas2005,Gerbier2010}, both abelian and
non-abelian.  Recently Spielman and coworkers\cite{Lin2009A,Lin2009B}
used Raman coupling between hyperfine states to produce synthetic
gauge fields.  They studied Bose condensates of $^{87}$Rb atoms
and investigated the punching in of vortices when a U(1) gauge field
corresponding to a magnetic field is tuned. Depending on the
degeneracy of the lowest Raman coupled states, one can also generate
non-abelian gauge fields. The condensation of bosons in non-abelian
fields have been investigated.\cite{Ho2010,Wang2010} 

These developments provide us the motivation to study {\em fermions}
in non-abelian gauge fields. The simplest in this class is the case of
spin-$\half$ particles coupled to an SU(2) gauge field. Study of such
systems within the cold atoms context will enable experimental realization and understanding of fermionic Hamiltonians with spin orbit
interactions that can lead to interesting topological phases of
matter.\cite{Qi2010,Hasan2010}

Readers who wish to obtain a qualitative understanding of our work may
read Sec.~\ref{sec:statement} where we state our problem and summarize
our results, followed by Sec.~\ref{sec:outlook} which discusses the
significance of these results. Sec.~\ref{sec:twobody} contains details
of our calculations, and Sec.~\ref{sec:qualitative} provides a
qualitative discussion of the physics of our results.

\begin{figure*}
\centerline{
\includegraphics[height=6cm,trim=0 50 0 35,clip]{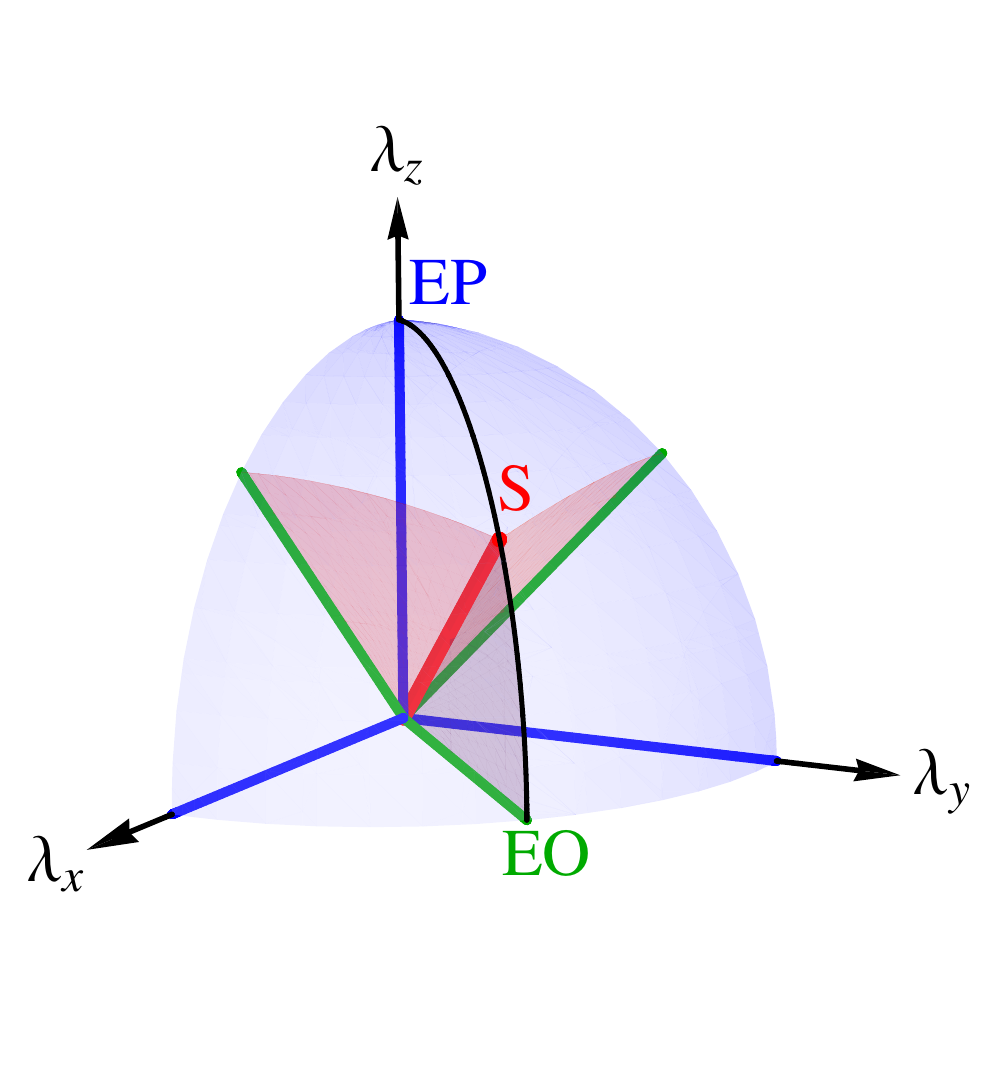}
\includegraphics[height=6cm]{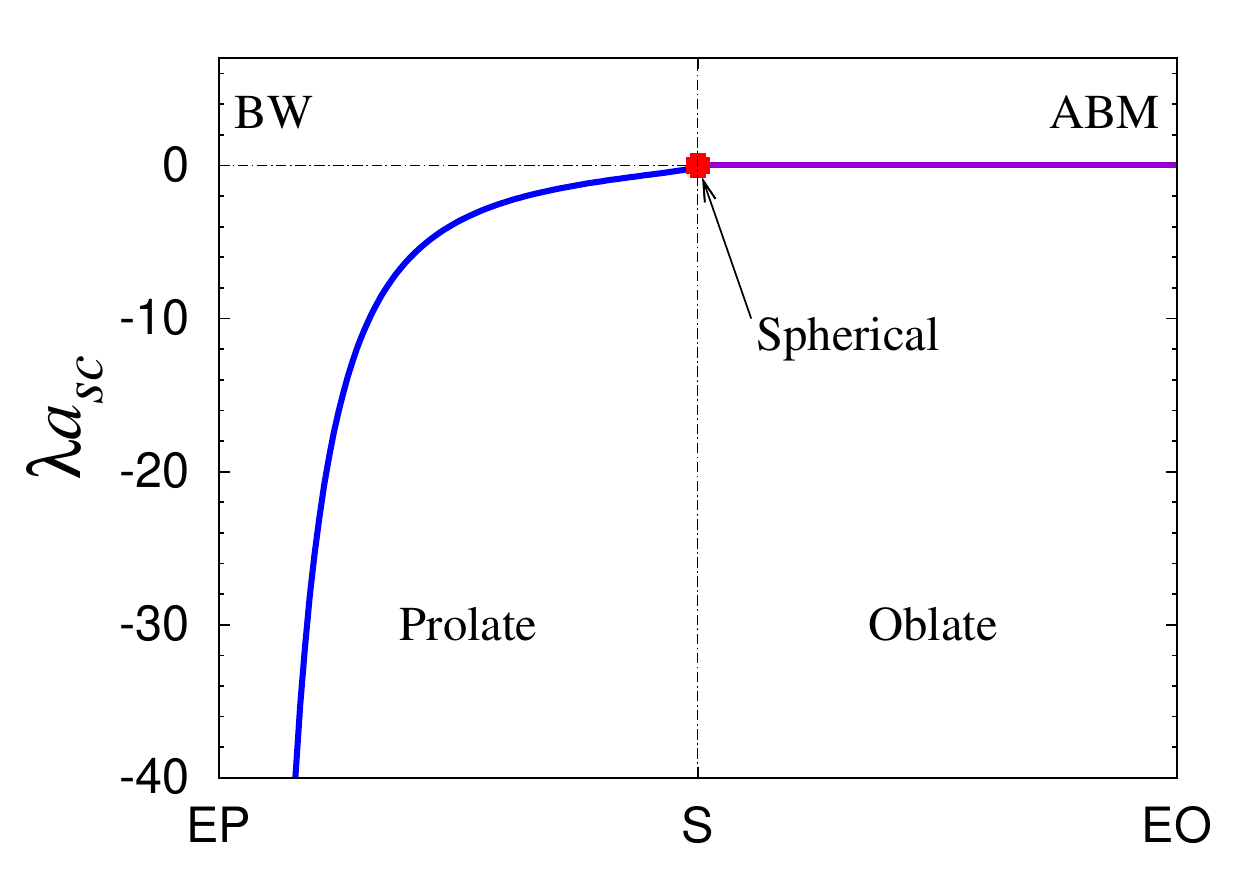}}
\centerline{(a)~~~~~~~~~~~~~~~~~~~~~~~~~~~~~~~~~~~~~~~~~~~~~~~~~~~~~~~~~~~~~~~~~~~~~~~~(b)}
\caption{ (color online) {\bf (a)} Two-body phase diagram in the gauge-field configuration (GFC) space described in \eqref{eqn:Rashba}. Lines in dark blue (along the axes) correspond to extreme prolate (EP) GFCs, dark green (along $45^\circ$ to the axes on the coordinate planes) correspond to extreme oblate (EO) GFCs, and the red line (along the body diagonal) corresponds to spherical (S) GFC. The wings in violet hue correspond to oblate GFCs. For the oblate, including EO GFCs, and the spherical (S) GFC a bound state is obtained for any scattering length i.e., $\asc=0^-$.  Regions with a blue hue, including the axes (EP), correspond to GFCs that require a non-zero critical scattering length for the formation of a bound state. {\bf (b)} The critical scattering length along the path EP-S-EO as shown in (a). For the EP GFCs, the symmetry of the bound state wave function corresponds to an extended Balian-Werthamer (BW) state with a biaxial nematic spin order, while that in the EO state corresponds to an extended Anderson-Brinkman-Morel (ABM) state with a uniaxial nematic spin order. The state evolves smoothly from a biaxial nematic to a uniaxial nematic passing through the spherical configuration (S) where the bound state is rotationally symmetric.}
\label{fig:PhasePlot}
\end{figure*}

\section{Statement of the Problem and Summary of Results}\label{sec:statement}

We consider spin-$\half$ fermions moving in 3D continuum in a non-abelian gauge field.
The simplest realization of this is described by the Hamiltonian
\bea
{\cal H}_{GF} = \int \D{^3 \br} \,\Psi^\dagger(\br)   \left[ \frac{1}{2}(p_i \bOne - A^\mu_i \btau^\mu) (p_i \bOne - A_i^\nu \btau^\nu) \right] \Psi(\br) \non \\
\eea
where $\Psi(\br) = \{\psi_\sigma(\br) \}, \sigma= \uparrow, \downarrow$ is
a two component spinor field (spin quantization along $z$-axis), $p_i$
is the momentum operator ($i = x,y,z$), $\bOne$ is the SU(2) identity,
$\btau^\mu$ are Pauli spin operators ($\mu=x,y,z$), $A_i^\mu$ describe
a {\em uniform} gauge field. We work with units where the mass of the
fermions and $\hslash$ are unity. Indeed even a uniform non-Abelian
field leads to interesting physics\cite{Ho2010}, an example of which
we demonstrate in this paper.

Motivated by the recent experiments mentioned above, we consider non-Abelian gauge fields of the type $A_i^\mu = \lambda_i \delta_i^\mu $ which leads to a generalized Rashba  Hamiltonian describing an anisotropic spin orbit interaction
\begin{equation}\label{eqn:Rashba}
\begin{split}
{\cal H}_{R} & =  \int \D{^3\br} \Psi^\dagger(\br) \left(\frac{\bp^2}{2} \bOne - \bp_\lambda \cdot \btau \right)\Psi(\br),  \\
\bp_\lambda & = \lambda_x p_x \be_x + \lambda_y p_y \be_y + \lambda_z p_z  \be_z.
\end{split} 
\end{equation}
Here an inconsequential constant term has been dropped.
  The gauge coupling strength is $\lambda= \sqrt{\lambda_x^2 + \lambda_y^2 + \lambda_z^2}$, and the vector $\blam
\equiv \lambda \hat{\blam} = \lambda_x \be_x + \lambda_y \be_y + \lambda_z \be_z$ defines a gauge field configuration (GFC).

We now describe the interaction between fermions by a contact
attraction model in the singlet channel\cite{Pethick2004}
\begin{equation}\label{eqn:contact_interaction}
{\cal H}_\upsilon = \frac{\upsilon}{2} \int \D{^3r} S^\dagger(\br) S(\br)
\end{equation}
where $S^{\dagger}(\br)$ is the singlet creation operator, and $\upsilon$ is the bare contact interaction. The theory described by the Hamiltonian ${\cal H} = {\cal H}_{R} + {\cal H}_{\upsilon}$ requires an ultraviolet momentum cut-off $\Lambda$. The bare contact interaction parameter $\upsilon$ is $\Lambda$-dependent and satisfies the regularization relation $\frac{1}{\upsilon} + \Lambda = \frac{1}{4 \pi a_s}$, where $a_s$ is the $s$-wave scattering length in the {\em absence} of the gauge field (``free vacuum''). It is well known\cite{Taylor2006} that for a pair of spin-$\half$ fermions in free vacuum, there is no bound state when $a_s < 0$ (conventionally called the ``BCS side''), and a bound state develops when $1/a_s = 0$ (``resonance''), and for $a_s \ge 0$ (``BEC side''), the binding energy $E_b = \frac{1}{a_s^2}$. This result embodies the fact that a critical attraction, characterized by the critical scattering length $\asc$, is needed to obtain a two-body bound state in the 3D free
vacuum where $1/\asc = 0$.

In this paper we address the question of how a uniform gauge field described by \eqref{eqn:Rashba} affects the nature of the bound state of two fermions interacting via \eqref{eqn:contact_interaction}. To this end, we obtain the ``phase-diagram'' of the two-fermion problem in the GFC space described by the parameters ${\lambda_x,\lambda_y,\lambda_z}$ of \eqref{eqn:Rashba}, by studying the bound state as a function of the free vacuum scattering length $a_s$.

GFCs can be conveniently classified as being prolate when two of
the $\lambda$s are equal and {\em smaller} than the third, oblate
when two of the $\lambda$s are equal and {\em larger} than the third,
spherical (S), when all three $\lambda$s are equal, and generic
when no two $\lambda$s are equal.  Our main findings are summarized in
\Fig{fig:PhasePlot}. We show that for prolate and generic GFCs, the critical scattering length $\asc$ required for the bound
state formation is {\em negative} \ie shifts to the BCS
side. However, for oblate and spherical GFCs $\asc$ vanishes, \ie
there is a bound state for {\em any} scattering length (see
\Fig{fig:PhasePlot}(b)). The key difference between the oblate and
spherical cases is the size of the binding energy of the bound
state. In the deep BCS side, for oblate gauge fields, the binding
energy depends exponentially on $\as$ and $\lambda$, while for
spherical gauge fields, an algebraic dependence is
obtained. Evidently, these results of the two-body problem suggest
that many body physics of fermions, in particular, the crossover from
the BCS regime to the BEC regime will be spectacularly affected by the
presence of a non-abelian gauge field. Moreover, our results below indicate
that the superfluid obtained at low temperatures will also
have additional spin nematic order induced by the gauge field.

\section{Two-Body Problem in presence of Non-Abelian Gauge Fields}\label{sec:twobody}

 For any GFC, the single particle states are described by the quantum numbers of momentum $\bk$ and helicity $\alpha$ (which takes on values $\pm$):
\bea
\ket{\bk\alpha} = \ket{\bk} \otimes \ket{\alpha \hat{\bk}_\lambda}
\eea
where $\ket{\bk}$ is the usual plane-wave state, and $\ket{\alpha \hat{\bk}_\lambda}$ is the spin coherent state in the direction $\alpha \hat{\bk}_\lambda$, with $\bk_\lambda$ defined analogous to $\bp_\lambda$ of \eqref{eqn:Rashba}. The two helicity bands disperse as
\begin{equation} \label{eqn:rashba_dispersion}
\varepsilon_{\bk\alpha} = \frac{k^2}{2} - \alpha |\bk_\lambda|
\end{equation}

 The full two body Hamiltonian ${\cal H}$ generically has only two symmetries -- global translation and time reversal. Therefore, the only good quantum number is the center of mass momentum of the two particles. We shall focus attention on states with zero center of mass momentum, and perform a $T$-matrix analysis\cite{Taylor2006} in the relative-momentum and helicity bases. The components of the $T$ matrix have the matrix structure $T_{\beta \beta'}(\omega)$, where $\omega$ is the energy, and $\beta, \beta'$ run over $\left(++,+-,-+,--\right)$, the helicity indices of the two fermions.  Since the interaction is only in the singlet channel, it follows that components with indices $(+-,-+)$ vanish.
This analysis, along with the regularization discussed earlier,  readily provides the condition for bound-state formation
\begin{equation}
\frac{1}{4 \pi \as} =  \frac{1}{2V} \sum_{\bk \alpha} \left( \frac{1}{E - 2\varepsilon_{\bk \alpha}} + \frac{1}{k^2} \right). \label{eqn:secular}
\end{equation}
where $V$ is the volume of space under consideration. Isolated poles of the $T$ matrix, which correspond to bound states, are obtained by finding the roots $E$ of \prn{eqn:secular}. We shall now present results for particular GFCs including the nature of the bound-state wave functions.

\subsection{Extreme prolate (EP)}
 In EP GFCs,  two of the gauge couplings vanish (say $\lambda_x= \lambda_y = 0$) while only one is nonzero ($\lambda_z = \lambda$). Such configurations correspond to the axes marked in blue (along the axes) in \Fig{fig:PhasePlot}(a). These GFCs  possess, in addition to translation and time reversal,  spatial and spin rotation symmetries about the $z$ axis.
The one particle dispersion \eqref{eqn:rashba_dispersion}, for this case, provides the scattering threshold $E_{th} = -\lambda^2$. Defining the binding energy $E_b = - (E - E_{th})$, we find from the solution of \eqref{eqn:secular} that a bound state appears only for positive scattering lengths (\Fig{fig:Eb}), with\bea
E_b &=& \frac{1}{\as^2} , \;\;\;\;\; \as >0.
\eea
The critical scattering length corresponds to resonance \ie $1/\asc = 0$. 

These results for $E_b$ and $\asc$ are identical to those of the two-body problem in free vacuum. There is, however, a crucial difference. The wave function of the bound state in the absence of the gauge field is a spin singlet. In an extreme prolate gauge field, the bound state wave function has two pieces,
\bea
\Psi_b &\propto& \psi_s(\br) \ket{\uparrow \downarrow - \downarrow \uparrow} + \psi_a(\br) \ket{\uparrow \downarrow + \downarrow \uparrow},
\eea
 where $\psi_s(\br) = \sum_{\bk \alpha} \frac{\cos{\bk \cdot \br}}{2\varepsilon_{\bk \alpha}-E} $   and $\psi_a(\br) =  \sum_{\bk \alpha} \frac{\alpha \sin{\bk \cdot \br}}{2\varepsilon_{\bk \alpha}-E} $ are, respectively, symmetric and anti-symmetric functions of the relative coordinate $\br$.  The first piece is a spin singlet, while the second piece (which vanishes when $\lambda \rightarrow 0$), has a triplet spin wave function. This wave function corresponds to an extended BW state\cite{Leggett2006} of the B-phase of $^3$He with an additional singlet piece. This state has a spin-nematic order\cite{Podolsky2005} corresponding to a {\em biaxial nematic}, consistent with the symmetries of the Hamiltonian.

\subsection{Extreme oblate (EO)} These configurations have one of the gauge couplings equal to zero, and the other two equal and non-zero, and are marked by green lines (lines at $45^\circ$ to the axes on the coordinate planes) in \Fig{fig:PhasePlot}. We consider the case with $\lambda_x = \lambda_y = \frac{\lambda}{\sqrt{2}}, \lambda_z = 0$. For these GFCs, we have in addition to translation and time reversal, a symmetry of global (spatial + spin) rotation about the $z$ axis generated by $J_z = L_z + \half \tau_z$, where $L_z$ is the $z$ component of the orbital angular momentum operator. 

The secular equation \eqref{eqn:secular}, in this case, reduces to
\begin{equation}
\frac{\sqrt{2}}{\lambda a_s} = \sqrt{1+\frac{2{E}_b}{\lambda^2}} - \log{\left(1+\sqrt{1+\frac{2{E}_b}{\lambda^2}}\right)} - \frac{1}{2}\log{\left(\frac{\lambda^2}{2 E_b}\right)}. \non
\end{equation}
The results presented for the EO case in the published version of this paper (Phys.~Rev.~B \textbf{83}, 094515 (2011)) is based on the analysis of the above equation. The analysis, however, was performed with an erroneous factor of 2 in the last term: $-\frac{1}{2} \log(\frac{\lambda^2}{E_b})$ was used instead of the correct term $-\frac{1}{2} \log(\frac{\lambda^2}{2 E_b})$. The results shown below (including \Fig{fig:Eb}) are the correct results for the EO GFCs.  We note that no qualitative conclusions and physics discussed in the published version  is altered by this error, and in addition, the results presented in the published version for all other GFCs are correct.

The solution of \eqref{eqn:secular} provides an interesting result: there is a bound state for {\em any} scattering length, negative or positive, \ie $\asc = 0^-$. For small negative $\as$ (BCS regime), we obtain the binding energy (referred to the scattering threshold $E_{th} = -\frac{\lambda^2}{2}$) as
\bea
\frac{E_b}{\lambda^2} &=& \sum_{n=0}^{\infty} (-1)^n a_n \left(\frac{2}{e^2} e^{\frac{-2\sqrt{2}}{\lambda |a_s|}}\right)^{n+1} \approx \frac{2}{e^2} e^{\frac{-2\sqrt{2}}{\lambda |a_s|}}
\eea
where $a_n$s are positive rationals tending asymptotically to $(4e)^{\frac{n}{2}}$; $a_0 = 1; a_1 = 1; a_2 = \frac{7}{4}; a_3 = \frac{23}{6}$. Thus the binding energy is exponentially small for small negative $\as$. For small positive $\as$ (BEC regime), we have
\bea
\frac{E_b}{\lambda^2} & = & \frac{1}{(\lambda a_s)^2} + \frac{1}{2} - \frac{(\lambda a_s)^2}{12} + \dots
\eea
which recovers the binding energy of $1/\as^2$ in the limit of free vacuum ($\lambda \rightarrow 0$). When $1/a_s =0$ (resonance), the binding energy is determined solely by $\lambda$; we obtain, near resonance,
\bea
\frac{E_b}{\lambda^2} &=& {\cal C} + \frac{2\sqrt{2}{\cal C}}{\sqrt{1+2{\cal C}}} \frac{1}{\lambda a_s} + \frac{4{\cal C}(1+{\cal C})}{(1+2{\cal C})^2} \frac{1}{(\lambda a_s)^2} + \dots
\eea
where ${\cal C} \approx 0.2196$. The full evolution of the bound state energy as a function of the scattering length $\as$ is shown in \Fig{fig:Eb}.

The bound-state wave function, again, has two pieces
\bea
\Psi_b & \propto & \psi_s(\br) \ket{\uparrow \downarrow - \downarrow \uparrow} +   \psi_a(\br) \ket{\uparrow \uparrow}  + \psi_a^*(\br) \ket{\downarrow \downarrow}
\eea 
where, $\psi_s(\br) = -\sum_{\bk \alpha} \frac{\cos{\bk \cdot \br}}{2\varepsilon_{\bk \alpha}-E}$, and $\psi_a(\br) =  i \sum_{\bk \alpha} \frac{\alpha e^{-i \phi_{\bk}} \sin{\bk \cdot \br}}{2\varepsilon_{\bk \alpha}-E}$,  $\phi_{\bk}$ is the angle made by $ k_x \be_x + k_y \be_y$ with the $x$ axis. The first piece is  orbitally symmetric $(\psi_s(\br))$ spin singlet  (the first term), and the second piece (next two terms) consists of an antisymmetric orbital wave function $(\psi_a(\br))$ and a spin structure corresponding to that of the ABM state\cite{Leggett2006} in the A-phase of $^3$He. This state has {\em uniaxial nematic} order.

\subsection{ Spherical (S)} 
\begin{figure}
\includegraphics[width=\myfigwidth]{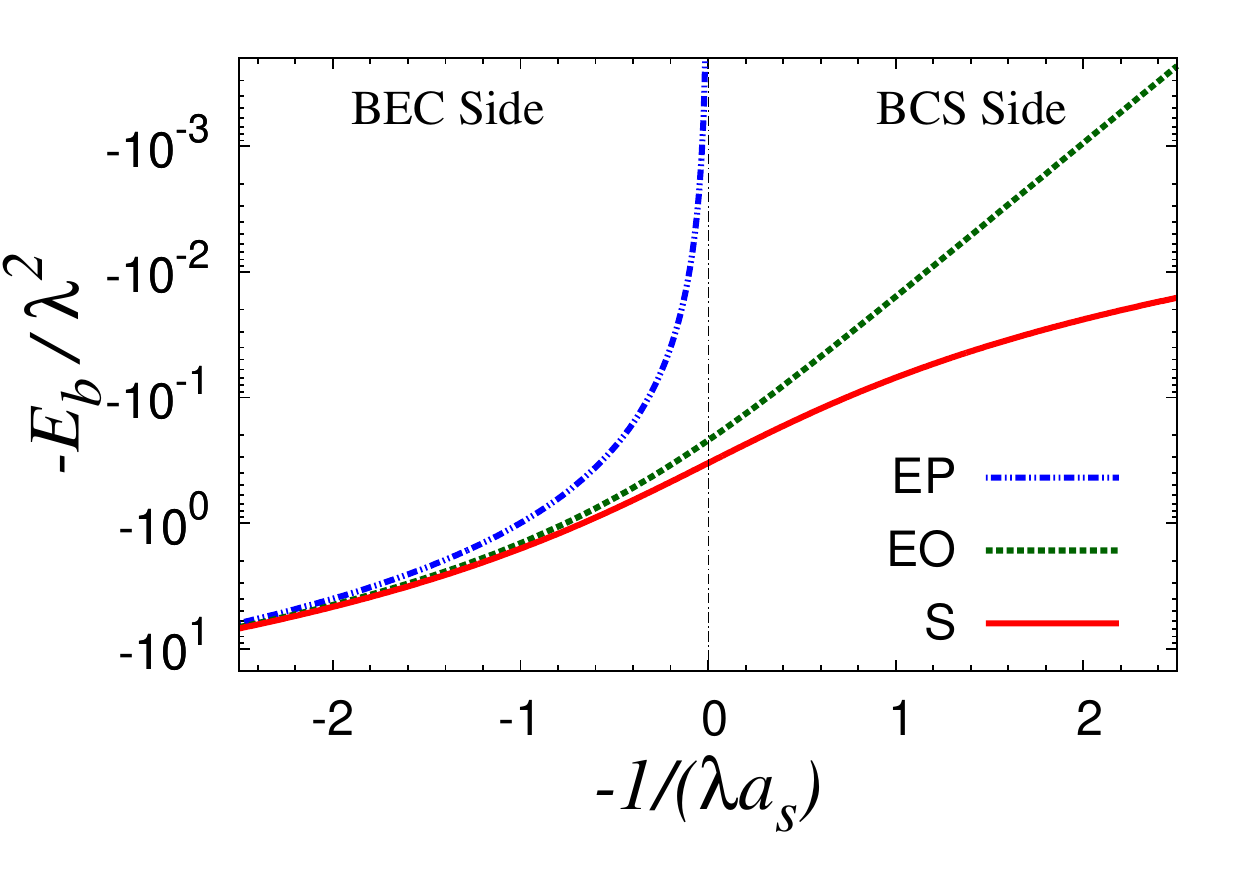}
\caption{(color online) Binding energy $E_b$ as a function of the scattering length $\as$ for extreme prolate (EP), extreme oblate (EO) and spherical (S) gauge field configurations. In the EP case, a bound state is obtained only for $1/\as \ge 0$, while for the other two it is obtained for every $\as$. In the BCS side, $E_b$ depends exponentially on $\lambda \as$ for the EO configurations, while it is a power law in the spherical case.}
\label{fig:Eb}
\end{figure}

This most symmetric GFC is characterized by $\lambda_x = \lambda_y = \lambda_z = \frac{\lambda}{\sqrt{3}}$ and marked by the red line (along the body diagonal) in \Fig{fig:PhasePlot}(a). Apart from translation and time reversal, this GFC has global rotational symmetries about all three axes generated by $J_i = L_i + \half \tau_i$. Again, we find that a two-body bound state appears for {\em any} scattering length, \ie $\asc =0^-$. 
Also, we obtain a closed form expression for the binding energy (referred to the scattering threshold $E_{th} = -\frac{\lambda^2}{3}$) for any scattering length 
\bea
E_b &=& \frac{1}{4} \left(\frac{1}{a_s} + \sqrt{\frac{1}{a_s^2} + \frac{4\lambda^2}{3}}\right)^2
\eea
An interesting aspect of this result is that, for a small negative scattering length (BCS side), the bound state energy depends {\em algebraically} on $\as$ and $\lambda$,
\bea
\frac{E_b}{\lambda^2} \approx \left(\frac{\lambda a_s}{3}\right)^2
\eea
\ie a deeper bound state than the EP case is obtained  (see  \Fig{fig:Eb}) in this case. For small positive $\as$, the leading term in the binding energy is that in the free vacuum. The bound state is a $J$ singlet and has the wave function
\bea
\Psi_b(\br) & \propto & \frac{e^{-\sqrt{E_b}r}}{r}\left( \frac{\lambda}{\sqrt{3 E_b}}\sin{\frac{\lambda r}{\sqrt{3}}} + \cos{\frac{\lambda r}{\sqrt{3}}} \right)\ket{\uparrow \downarrow - \downarrow \uparrow} \non\\
&& \!\!\!\!\!\!\!\!\!\!\!\!\!\!\!\!\!\!\!\!\!\!\!\!\!\!\!\!\!\! + i \left(\left(\sqrt{E_b} + \frac{1}{r} \right) \sin{\frac{\lambda r}{\sqrt{3}}} - \frac{\lambda}{\sqrt{3}} \cos{\frac{\lambda r}{\sqrt{3}}} \right) \frac{e^{-\sqrt{E_b}r}}{\sqrt{E_b}r} \ket{\uparrow \downarrow + \downarrow \uparrow}_{\hat{\br}} \non
\eea
where the subscript $\hat{\br}$ on the second term indicates that the spin quantization axis is along $\hat{\br}$. The wave function is made of two pieces. The first piece corresponds to a $J=0$ state constructed out of $L=0$ orbital state and a spin singlet, while the second piece is a $J=0$ state obtained by fusing an $L=1$ orbital state and a spin triplet state.   Furthermore, orbital wave functions of both pieces are non-monotonic, \ie they have spatial oscillations. This owes to the
existence of two length scales  determined by $E_b$ and $\lambda$. While the former dictates the exponential decay of the wave function, the latter determines the period of its spatial oscillation. This observation also applies to the wave functions discussed above for the extreme prolate and extreme oblate cases.

It is interesting to note how the bound state evolves as we go from the EP to the EO GFC along the path in GFC space indicated in  \Fig{fig:PhasePlot}(a). The prolate side of the path which has a bi-axial nematic order, is separated from the oblate side with a uniaxial nematic order by the spherical configuration. For the spherical configuration the bound state is fully (spatial + spin) rotationally  symmetric.

\subsection{Generic GFC}

 For a generic GFC, the critical scattering length for the formation of a bound state can be expressed as
\bea
\lambda \asc = \cF(\hat{\blam})
\eea
where $\cF$ is a dimensionless number-valued function of the unit vector
$\hat{\blam}$. The function $\cF$ has to be obtained numerically. We
find that $\cF$ is a non-positive function, i.e., for a generic GFC,
the critical scattering length is {\it negative}, \ie on the BCS
side of the resonance. In other words, the strength of the critical
attraction required to produce a two body bound state is
reduced by the presence of a generic gauge
field. \Fig{fig:PhasePlot}(b) shows the evolution of $\asc$ along the
great circle connecting the EP state to EO state for a fixed gauge
coupling, illustrating that  prolate GFC has a negative $\asc$,
while any oblate  GFC has a vanishing $\asc$. In summary, the two-body boundstate appears at resonance $(1/\asc=0)$
for EP GFCs marked by the blue lines  (along the axes) in
\Fig{fig:PhasePlot}(a). For spherical (S) GFC marked
by the red line (along the body diagonal) and for oblate GFCs
marked by the planes bounded by the green (including EO) lines (along $45^\circ$ to the axes on the coordinate planes) and the red line (along the body diagonal),
$\asc$ vanishes, \ie any attractive interaction, however small,
will force a bound state for the two body problem.

\section{Qualitative Discussion}\label{sec:qualitative}

\begin{figure}
\includegraphics[width=\myfigwidth]{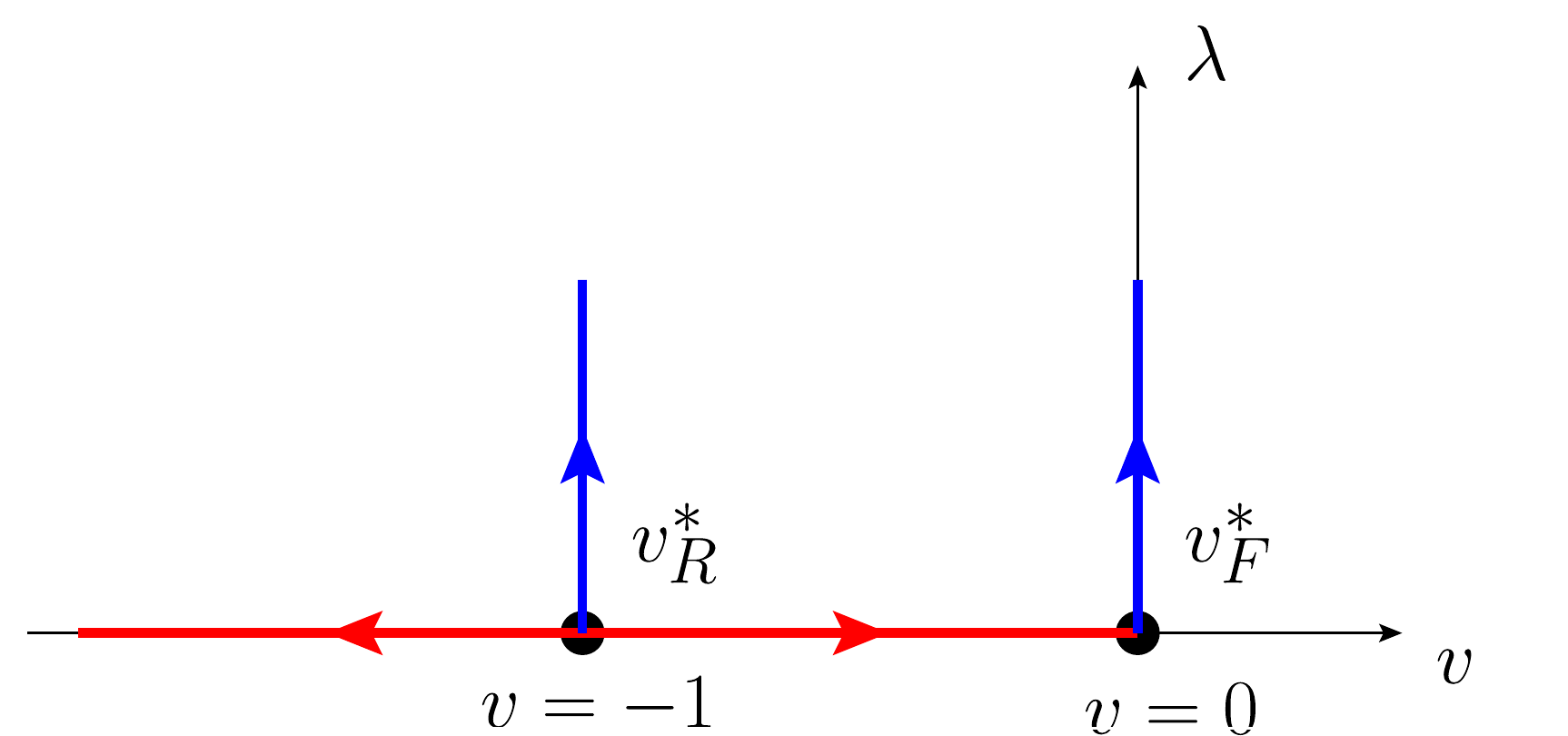}
\caption{(color online) Renormalization group flow diagram (schematic) of the two body problem. The point $\upsilon^*_F$ corresponds to the free fixed point and $\upsilon^*_R$ to resonance in vacuum.\cite{Sauli2006,Nikolic2007} The non-abelian gauge field is a relevant operator at these fixed points as indicated. }
\label{fig:lambdaRG}
\end{figure}

We now discuss the physics behind these results. In the free vacuum,
a renormalization group analysis of the field theory of the two body
problem with the contact interaction\cite{Sauli2006,Nikolic2007} provides
two fixed points. The first is a stable one at $\upsilon^*_F = 0$
describing two free fermions, and the second, an unstable one
$\upsilon^*_R= -1$ (in suitably chosen units) corresponding to the
resonance; see \fig{fig:lambdaRG}.  In free vacuum, a contact interaction parameter $\upsilon$
near $\upsilon^*_F$ flows toward $\upsilon^*_F$ and hence has similar
physics as two free fermions. This corresponds to the fact that
sufficiently strong attraction is required ($\upsilon < -1$) to produce
a bound state. Consider now a situation with a $\upsilon$ near
$\upsilon_F^*$ and a non-vanishing spherical gauge field with a
coupling strength $\lambda$. We see immediately that Rashba term
described by the coupling $\lambda$ is a relevant operator and the
flow takes the system {\it away} from the free fixed point (see \Fig{fig:lambdaRG}) suggesting
that even a small $\lambda$ has a drastic effect on a system near the free fixed
point.

A deeper understanding  can be obtained by considering the density of states $g(\varepsilon)$ of a single fermion moving in a gauge field, since it determines the density of states of two
non-interacting fermions with zero center of mass momentum. One can be easily obtain analytical expressions  for the density of states for the high symmetry GFCs. The gist of those formulae is that near the scattering threshold, for the high symmetry GFCs discussed above
\begin{equation}
g(\varepsilon)  \sim  
\begin{cases}
\sqrt{\varepsilon}  & \text{for} \;\; \mbox{EP} \\
\lambda \mbox{(constant)} & \text{for} \;\; \mbox{EO} \\
\frac{1}{\sqrt{\varepsilon}} & \text{for} \;\; \mbox{S}
\end{cases}. 
\end{equation}
In all three cases $g(\varepsilon) \rightarrow \sqrt{\varepsilon}$ as $\varepsilon \rightarrow \infty$. It is therefore clear that the infrared behavior of the density of states is behind the results presented hitherto. This motivates us to construct a model with density of states given by
\begin{equation}
g(\varepsilon) = \begin{cases}
   \frac{\sqrt{2 \varepsilon_0}  }{\pi^2} \left(\frac{\varepsilon}{\varepsilon_0}\right)^{\gamma} \Theta(\varepsilon) & \text{if } \varepsilon < \varepsilon_0,\\
  \frac{\sqrt{2 \varepsilon}}{\pi^2} & \text{if } \varepsilon \geq \varepsilon_0.
\end{cases}
\end{equation}
where $\Theta$ is the unit step function, $\gamma$ is an exponent that determines the infrared behavior of $g$, and $\varepsilon_0$  is an energy scale (crudely equal to $\lambda^2$) at which the density of states is restored to  that in the free vacuum. Note that $\gamma = \half, 0, -\half$ qualitatively reproduces, respectively, the density of states corresponding to EP, EO and S GFCs. We can readily calculate the critical scattering length that obtains a bound state as
\bea
\sqrt{2 \varepsilon_0} \asc =  \frac{\pi \gamma}{2\gamma -1} \Theta(\gamma)
\eea
We see immediately that $\asc$ vanishes whenever $\gamma$ is non-positive as is the case for the EO and S configurations. For the EP configuration, the critical scattering length  $\asc \rightarrow -\infty $  consistent with the results presented earlier. For a generic GFC, the infrared density of states has a narrow $\sqrt{\varepsilon}$-regime, followed by a regime with nearly constant density of states -- this can be modeled in this simple picture using a $\gamma$ that satisfies $0 < \gamma < \half$, the precise value of $\gamma$ being dependent on the direction $\hat{\blam}$ in the GFC space. We find that $\asc$ is negative, again, consistent with our calculations. 

  This simple analysis allows us to uncover the physics behind the phase diagram of \Fig{fig:PhasePlot}. Highly symmetric GFCs drastically modify the low energy density of states owing to the degeneracies induced in the resulting one particle levels. It is in this sense that the Rashba term is a relevant operator as mentioned in the discussion above. For highly symmetric GFCs, the enhanced density of states at low energies strongly promotes bound state formation in the presence of an attractive interaction.

  The particular type of nematic spin symmetry in the bound state arises so as to optimize the kinetic energy. The spin-orbit interaction mixes the singlet and triplet sectors of the two particle system, and the particular nematic symmetry obtained in the bound state enables the orbital wave function to sufficiently ``sample'' the attractive interaction at a minimal cost in kinetic energy.

\section{Experimental Directions and Outlook}\label{sec:outlook}

 Our predictions can be readily tested by experiment. A clear signature of the bound state formation can be obtained from the measurement of energy\cite{Ho2004} of the gas at high temperatures. As suggested by \citebyname{Ho2004}, a large value of the second virial coefficient which characterizes the interaction energy, is obtained for an interacting Fermi gas in free vacuum near resonance. Our results suggest that in presence of a generic gauge field, such large corrections to energy can be observed on the {\it BCS side} \ie for negative scattering lengths near $\asc$ at the onset of the bound state.  The quantitative value of $\asc$ will be affected also by the attraction in the triplet channel, but our predictions can be tested qualitatively.

  Our results suggest that the BCS-BEC crossover in the presence of a non-Abelian gauge field will be drastically altered and in particular the ``crossover regime''  should shift  to the BCS side for a generic GFC. There are many other novel effects of the non-Abelian gauge field in the many body context such as transition in the topology of the Fermi surface with increasing filling\cite{Vyasanakere2010}. Moreover, the two-body bound-state wave functions provide a clue to the nature of the Cooper pair wave functions. Clearly, this will lead to superfluild states with interesting pairing wave functions (such as the extended BW and ABM states found here with associated nematic orders) and concomitant excitations.

The simple model we have presented in Sec.~\ref{sec:qualitative} suggests that our conclusions will also apply to systems with a larger gauge group (such as SU($N$)). Investigations along these lines should lead to interesting new possibilities with cold atom systems.

\subsection*{Acknowledgements}

Support of this work by CSIR, India (JV, via a JRF grant), DST, India (VBS, via a Ramanujan grant) and DAE, India (VBS, via an SRC grant) is gratefully acknowledged. VBS thanks Tin-Lun (Jason) Ho for many illuminating comments and fruitful suggestions, and Shizhong Zhang for discussions. VBS expresses his gratitude to Hui Zhai for hosting a  visit to Tshinghua University.

\bibliography{ref}

\end{document}